\documentclass[preprint,12pt]{elsarticle}

\usepackage{amssymb}
\usepackage{amsmath}
\usepackage{booktabs}

\usepackage{booktabs}
\usepackage{hyperref} 
\usepackage{subcaption}

\journal{Journal}

\begin{document}

\begin{frontmatter}



\title{A Comparison of Human and Machine Learning Errors in Face Recognition}



\author[inst1]{Marina Estévez-Almenzar}
\affiliation[inst1]{organization={Universitat Pompeu Fabra},
            city={Barcelona},
            country={Spain}}

\author[inst1,inst2]{Ricardo Baeza-Yates}
\affiliation[inst2]{organization={EAI, Northeastern University},
            city={Silicon Valley},
            state={California},
            country={USA}}

\author[inst1,inst3]{Carlos Castillo}
\affiliation[inst3]{organization={ICREA},
            city={Barcelona},
            country={Spain}}
\begin{abstract}
Machine learning applications in high-stakes scenarios should always operate under human oversight.
Developing an optimal combination of human and machine intelligence requires an understanding of their complementarities, particularly regarding the similarities and differences in the way they make mistakes.
We perform extensive experiments in the area of face recognition and compare two automated face recognition systems against human annotators through a demographically balanced user study. 
Our research uncovers important ways in which machine learning errors and human errors differ from each other, and suggests potential strategies in which human-machine collaboration can improve accuracy in face recognition.
%
\end{abstract}




\end{frontmatter}



\section{Introduction}
\label{introduction}

Decision support systems powered by machine learning (ML) are increasingly used in high-stakes scenarios including immigration, video-surveillance, healthcare, justice, and access to labor and education, among many others.
In these application domains, ML systems should not be autonomous, but rely on a human operator or expert, who should be responsible for the final decision.
%
An in-depth understanding of the dynamics of human-algorithm interactions is crucial for developing safe, trustworthy systems \cite{matias2023humans}.

%
%
%
%
Understanding the complementarities between human and machine intelligence is crucial.
%
%
In an ``ideal'' scenario, there is perfect complementarity: cases challenging for the ML system are easily handled by the human operator.
Conversely, the worst case is when there is total overlap: cases that are difficult or uncertain for the ML also lead to human errors.
In practice, we may find applications that are somewhere between these extremes, as our empirical findings demonstrate within the context of face recognition.

Among other areas that need exploration, little has been done to understand when human and machine errors are similar and when they are different. Analyzing these similarities and differences is particularly important because the presence of algorithmic errors influences well-known patterns of human-machine interaction, such as \emph{algorithmic aversion} \cite{madoc2020aversion,dietvorst2015algorithm}, a biased and overly negative human evaluation of an algorithm.
The question arises as to whether this aversion also varies depending on whether or not the errors presented by the model resemble those that a human agent might make.
If we are able to avoid this type of bias, there is another risk: \emph{automation bias,} an over-reliance on automated decision support mechanisms \cite{lyell2017automation}.
In this case we can ask an analogous question: Does a high similarity between human and machine errors influence the human agent's ability to judge the accuracy of the model? 

The main goal of this research is to compare ML errors and human errors. The results obtained from this comparative study serve as inputs for the development of straightforward yet impactful strategies to combine human and machine intelligence.
%

The use of the concept ``human error'' suggests an homogeneity that is almost non-existent in real life. Human perception varies from individual to individual, either due to variations in physiological structures or external influences such as culture.
It becomes even more complex when it comes to assessing human perception in distinguishing between other human identities, such as in the context of a face recognition task.
We tackle this complexity through a demographically diverse user study for face matching in which possible inter-individual differences and disagreements are considered. 
Proposing hybrid human-machine strategies in the field of face recognition is crucial. For instance, in an automated system integrated in a police surveillance scenario, interrogating or detaining someone just because a face recognition system has erroneously matched their face in a database of persons of interest deserves special attention in the current use of facial recognition technologies. In January 2020, Robert Julian-Borchak Williams became the first documented example in the U.S. of someone being wrongfully arrested based on a false hit produced by facial recognition technology. \footnote{\href{https://www.nytimes.com/2020/06/24/technology/facial-recognition-arrest.html}{\em Wrongfully Accused by an Algorithm}, The New York Times, 24 June 2020}
Many more cases have been documented, and often there are racial biases.\footnote{\href{https://innocenceproject.org/when-artificial-intelligence-gets-it-wrong/}{\em When AI Gets It Wrong}. Innocence Project, 19 September 2023.}

Our main findings for the face recognition task we study are:
(1) humans rarely produce false positives;
(2) the ML similarity score is a potential error predictor;
(3) humans find it easier to address mistakes made by an individual model compared to addressing shared errors between two models; and
(4) in face recognition the human perception of gender expression and ethnic appearance is determinant.
These findings provide a method for detecting potential errors in automated facial recognition, and help us find potential errors that a human annotator has a high chance of correcting.
Applying this approach in a practical setting enables us to develop an effective evaluation strategy that maximizes joint human-machine accuracy while controlling human annotation effort.
Unlike other approaches that strictly emphasize the enhancement of accuracy through algorithmic advancements, this work underscores not only the importance of incorporating the human factor in this race for accuracy maximization, but also the effectiveness of this approach.

The rest of the paper is organized as follows. In \S\ref{sec:related} we review related work, followed by our research questions and our methodology in \S\ref{sec:research} and \S\ref{sec:methods}, respectively. In \S\ref{sec:results} we present our results, while in \S\ref{sec:discussion} we discuss our results and present a human-computer collaboration strategy based on our findings. We also outline some of the limitations encountered in the development of this work, as well as possible future directions for it. We conclude in \S\ref{sec:conclusions}.

\section{Related Work}
\label{sec:related}

Even though automated facial recognition systems are not influenced by factors that affect human ability to match faces (e.g., time pressure \cite{fysh2017effects}, fatigue \cite{behrens2023fatigue}, processing capabilities in real-time \cite{curry2003capability}), they are affected by other factors. 

\subsection{Human and ML performance}

ML systems may outperform human annotators in tasks considered simple or moderately difficult, but they tend to struggle when faced with more complex conditions that mirror real-life scenarios. In challenging tasks, non-expert observers showed performance comparable to that of some facial recognition algorithms, while in some cases experts outperformed these algorithms \cite{white2015perceptual}. Some systems did not make accurate identifications, while humans exceeded random chance \cite{rice2013unaware}.
Throughout the years, numerous competitions have been conducted to assess human-algorithm performance in different face recognition tasks, opposing algorithmic accuracy to human accuracy, thus establishing a distinction between two solving agents that, instead of collaborating, compete. Phillips {\em et al.} \cite{phillips2014comparison} conducted a cross-modal study to evaluate the results of selected human-algorithm competitions in facial recognition. Their findings revealed that algorithms outperformed humans in the case of simple frontal static images, whereas humans demonstrated superiority in challenging static images and videos.
Rice {\em et al.} \cite{rice2013unaware} were interested in the specific cases where the facial recognition system fails, and investigated how humans performed in these cases. They documented instances where facial recognition algorithms did not achieve any successful matches, while humans outperformed random chance. White {\em et al.} \cite{white2015perceptual} observed that in forensic facial identification algorithms performed similarly to certain observers and were outperformed by experts. 

\subsection{Combining human and machine intelligence}
Researchers have seek to uncover how human and machine intelligence can be reliably combined.
\textit{Algorithm aversion} has been extensively studied \cite{dzindolet2002perceived,reich2023overcome}, and has been shown that it becomes particularly noticeable when users witness mistakes made by the algorithm \cite{dietvorst2015algorithm}. This aversion is reduced when the user has some level of control (even if little) over the prediction process \cite{dietvorst2018overcoming,roy2019automation}.

This situation illustrates that the successful integration of facial recognition systems in practical settings necessitates more than just technological progress. According to the EU AI Act \cite{EUAIAct}, the implementation of facial recognition should be proportionate and deployed only when strictly necessary. In \cite{negri2024framework}, Negri {\em et al.} present a framework aimed at determining whether a facial recognition intervention is appropriate for a particular usage scenario.
Other factors such as the application context, including the prospective end-users and the demographic characteristics of the population on which the system will operate, must be thoroughly taken into account. These approaches are closely connected to investigating human-centered ML techniques \cite{papenmeier2022accurate}, such as human oversight strategies for assessing and enhancing system outcomes \cite{hupont2022landscape,kyriakou2023humans}, as well as mechanisms for preserving the essential human element in decision-making in areas where safeguarding fundamental rights is particularly crucial \cite{koulu2020proceduralizing}. 

\subsection{Human factors in decision support}
Numerous studies have been conducted to explore methods for incorporating human factors into ML systems. Han {\em et al.} \cite{han2017hard} introduced an emotion detection model that leveraged inter-annotator agreement to provide a prediction that is more akin to human judgment. Their approach diverged from the conventional belief that a person's state can be simply classified into a \textit{hard} category or single value.
The idea of representing the human element using a continuous distribution is endorsed by Peterson {\em et al.} \cite{peterson2019human}, who introduced a novel image dataset that includes a comprehensive range of human annotations for each image. By representing human-like uncertainty, they achieved favorable results in terms of robustness and out-of-training-set performance, demonstrating that errors in human classification can be just as enlightening as accurate responses. 

Related to facial recognition technologies, Andrews {\em et al.} \cite{andrews2023view} raised doubts about the ability of categorical labels to capture the continuous spectrum of human phenotype diversity, particularly in the context of deducing delicate characteristics like social identities (as only a few facial recognition datasets include self-identified categories). They presented a dataset with human perception of face similarities that can be used to learn an embedding space aligned with human perception.
In relation to the distinctive features present in human perception, Makino et al. \cite{makino2022differences} examined the differences between deep neural networks (DNNs) and human perception in medical diagnosis, focusing on breast cancer screening. They discovered that DNNs utilize features that radiologists often ignore and are outside areas that they considered suspicious. This underscores the importance of incorporating domain knowledge into comparisons of human and machine perception to prevent erroneous outcomes.
In a similar vein, Huber {\em et al.} \cite{huber2022stating} proposed the propagation of model uncertainties to the final output to enhance transparency of facial recognition systems and offer a deeper understanding of the verification process. 
Additionally, Papenmeir {\em et al.} \cite{papenmeier2022accurate} conducted a study involving users and discovered that the perceived accuracy of a model was notably more reduced when users saw the model failing on a simple task compared to when it made mistakes on more challenging tasks. This research suggests that algorithm aversion may be impacted differently based on the type of model errors encountered.
%

While it is desirable to have greater consideration of human factors in the automated decision-making process, researchers have also studied the negative consequences of mimicking certain human biases \cite{hupont2019demogpairs}. The \textit{other-race effect} for face recognition (our ability to best recognize the identity of faces from our own race) has been observed in several human studies \cite{meissner2001thirty,feliciano2016shades}. Philips {\em et al.} \cite{phillips2011other} showed an other-race effect for the algorithms, concluding that their performance varies as a function of the demographic origin of the algorithm and the demographic contents of the test population. Similarly, motivated by this human bias, Flores-Saviaga {\em et al.} \cite{flores2023inclusive} propose an alternative interface to the classical human-in-the-loop interface and suggest that deriving the classification of facial image pairs as a function of annotators' race will improve the efficiency of the system. But, as they point out, such design decisions carry delicate ethical implications that underscore the importance of work along other lines.

There is a recent line of research investigating how human annotators can be effectively introduced into the loop so the algorithm can pass the final decision to the human when certain conditions are given \cite{hemmer2023learning,mozannar2023should,keswani2021towards}. These conditions are often related to the low confidence of an automated system, which can be used to determine what type of human-machine interaction is most appropriate in a hybrid system \cite{punzi2024ai}, as well as to distinguish which annotations flows should be adopted to make human-machine collaboration more efficient \cite{lee2022towards}. Combining decisions of systems and humans based on (weighted by) their perceived individual similarities has also been investigated \cite{phillips2018face}.

\bigskip
To the best of our knowledge, most of the efforts in integrating human factors into technology have mainly focused on encoding specific human traits and enhancing model performance — observing humans and refining models independently. Some more recent efforts have gone further, proposing novel techniques to combine human and machine performance. However, there is still a lack of understanding of the key differences of decision-makers, especially in contexts where the task involves a certain subjectivity. In a scenario where there is no longer only the final decision related to the task at hand, but also the decision as to which agent — human, algorithmic or combination of both — should make the final decision, it is important to know the strengths and weaknesses of both agents, which of these are shared and which diverge, and how these differences and similarities can be exploited. Here, we propose to study human and machine similarities and differences to capture and understand their complementary aspects so that this knowledge enables efficient and accountable human-machine interaction paradigms.
For this purpose, we establish an error-centred comparison of human and machine performance in solving a face matching task. Examining these distinctions and similarities is crucial for enhancing the effectiveness of human oversight of algorithms, and for gaining insights into integrating the human factor into decision-making processes. Additionally, we explore how gender and ethnicity can influence human errors, building on earlier findings that have identified their significance in tasks that involve facial recognition \cite{phillips2011other,wright2003own}.

\section{Research Questions}
\label{sec:research}

The main goal of this work is to study model errors in face recognition, comparing them with human errors. 
We also investigate how human conceptions of gender and ethnicity affect these errors.

Our experimental setting, described in detail in \S\ref{sec:methods}, is based on a number of face recognition tasks that are performed by two automated systems, as well as by human annotators hired through a crowdsourcing platform. These tasks consist of matching facial images: given a pair of facial images, determining whether they belong to the same individual or to two distinct individuals. Both the two automated models and the set of annotators performed this task independently.

\subsection{Error consistency} 
\label{subsec:consistency}
We would like to characterize similarities and differences between human errors and ML system errors.
%
%
To achieve this, first we need to determine if errors and successes are consistent, {\em i.e.}, if we can determine which are the subsets of face recognition tasks in which errors and successes are concentrated.

\begin{quote}
\textbf{RQ1a}
\textit{Are human annotators consistent when solving a face recognition task?}


\textbf{RQ1b}
\textit{Are ML systems consistent when solving a face recognition task?}
\end{quote}

If humans are consistent in their errors and successes, then we can define for a ML system in general, and for a face recognition system in our case, ``human-like'' errors as those machine errors that a human would also tend to make, and ``non-human-like'' errors as those machine errors that a human would not be likely to make.

\subsection{Error alignment} We want to uncover whether there are common difficulties between ML systems and human annotators.
We expect these common difficulties to manifest as incorrect human annotations on those face recognition tasks where the ML system erred. We also expect to obtain more incorrect annotations in cases where more than one ML system errs.
If human annotators and ML systems know whether they are likely to be making a mistake ({\em i.e.}, provide a low-confidence annotation), then we would like to test whether their confidence in annotation aligns. 
This would indicate that not only there are face recognition tasks that are likely to lead to errors by human annotators and ML systems, but that there are also tasks that are more challenging and elicit less certainty in both situations.
\begin{quote}
\textbf{RQ2a} \textit{Are human annotators more likely to make a mistake in a face recognition task if a ML system also gives an incorrect answer for that task, compared to tasks for which the system is correct?}

\textbf{RQ2b} \textit{Are human annotators even more likely to make a mistake if more than one ML system is incorrect?}

\textbf{RQ2c} \textit{Are human annotators' perception of similarity and ML computations of similarity correlated?}
\end{quote}

\subsection{The role of gender and ethnicity in errors} We define, as detailed in the next section, a false positive in face recognition as the incorrect identification of two images of different people as the same person.
False positives involving people of different gender and/or ethnicity are unlikely to be made by human annotators, as differences related to gender expression and ethnic appearance are determining factors in humans when establishing an identity judgment \cite{phillips2011other,wright2003own}.
%
%
In addition to errors, we would like to know whether cases where images are perceived to have different gender/ethnicity by human annotators lead to lower confidence by a ML system.

We remark that ``human perception of gender and ethnicity,'' refers to differences and similarities in terms of gender expression and ethnic appearance, and not in terms of gender identity or self-ascription to an ethnicity. 
%
As described in the next section, in some datasets labels are not provided by the photo subjects themselves, but are inferred through other means.
\begin{quote}
\textbf{RQ3a} \textit{Are ML errors on pairs of images labeled as depicting different gender expression, or eliciting different perceptions of ethnicity unlikely to be made by human annotators, to the extent that this can be used to characterize ``human-like'' and ``non-human-like'' errors?}

\textbf{RQ3b} \textit{Are human perception of similarity and/or ML similarity score correlated with human perception of gender and/or ethnicity similarity?}
\end{quote}

\subsection{Exploratory study of error-based human-machine collaboration} With the above questions we want to know if we can develop a strategy to optimise human-machine collaboration in the context of solving a face recognition task. The consistency raised in the first question would allow us to generalise in this context, while the study of the alignment between human and machine error patterns would allow us to detect the key points of complementarity for the development of a successful strategy.
\begin{quote}
\textbf{RQ4} \textit{Can we design a human-computer collaboration strategy based on the results obtained from this comparative study?}
\end{quote}

\section{Experimental Setup and Ethical Considerations}
\label{sec:methods}

In the next five subsections we detail the data used and the methodological aspects of this work. In the last subsection we detail some of the ethical aspects that have been taken into account in the development of this work.

\subsection{Datasets}

\subsubsection{Training data.} We used two pre-trained face recognition models.
Both were trained by their respective authors on \emph{MS-Celeb-1M} \cite{guo2016ms}, a dataset released by Microsoft in 2016. According to its authors, it was the largest publicly available face recognition dataset in the world. It contains about 10M images of nearly 100K people. 
After an investigation by Financial Times in 2019,\footnote{\href{https://www.ft.com/content/cf19b956-60a2-11e9-b285-3acd5d43599e}{\em Who’s using your face? The ugly truth about facial recognition} Financial Times, 18 September 2019} it was found that many of the people who appeared in the images were not asked for their consent, nor were they aware that their faces appeared in this database. Some time after this finding, without warning, 
Microsoft removed MS-Celeb-1M and its web page\footnote{\href{https://www.msceleb.org/}{https://www.msceleb.org/}} is currently offline.
Before its demise, the dataset was widely used and still exists in several forms, such as trained models. 
MS-Celeb-1M is fairly unbalanced demographically (see Table \ref{tab:datasets}).

\begin{table}[t]
\centering
\caption{Characteristics of MS-Celeb-1M \cite{guo2016ms,wang2019racial} and DemogPairs \cite{hupont2019demogpairs}.}
\label{tab:datasets}
\begin{tabular}{ccc} \toprule
 & ~~MS-Celeb-1M~~ & ~~DemogPairs~~  \\ \midrule
\# Images & 10M  & 10.8K  \\
\# People & 100K & 600 \\ \midrule
\% Female & $\approx$80\% & 50\% \\
\% Male   & $\approx$20\% & 50\% \\ \midrule
\% White  & 76.3\%        & 33.3\% \\
\% Black  & 14.5\%        & 33.3\% \\
\% Asian  & 6.6\%         & 33.3\% \\
\% Other  & 2.6\%         & - \\
\bottomrule
\end{tabular}
\end{table}

\subsubsection{Testing data.} We used \emph{DemogPairs} \cite{hupont2019demogpairs} as the evaluation dataset.
It contains 10,800 facial images corresponding to 600 people divided into 6 balanced demographic labeled folds: \{ female, male \} $\times$ \{ Black, Asian, White \}. 
These labels were manually annotated by their authors. We will use \textit{labels} when we refer to those from the original dataset, and \textit{annotations} for those obtained from our user study. 
Each demographic fold has 100 subjects, with 18 images per subject (see Table \ref{tab:datasets}). DemogPairs was created and released by its authors with the explicit objective of being used as a tool to test for demographic biases on face recognition models.

\subsection{Models}

The two face recognition models used in this work were IR50+ArcFace \cite{deng2019arcface} and LightCNN v4 \cite{wu2018light}, both trained, as explained before, over MS-Celeb-1M by the respective authors. We did not do any additional training or fine-tuning for this work. Both pre-trained models can be found in their original sources.\footnote{\href{https://github.com/ZhaoJ9014/face.evoLVe}{IR50+ArcFace pre-trained model on MS-Celeb-1M}, \href{https://github.com/AlfredXiangWu/LightCNN}{LightCNN v4 pre-trained model on MS-Celeb-1M}}

\textbf{IR50+ArcFace} is an extension of ResNet50 \cite{guo2016ms,he2016deep}, a residual network that has been extensively applied to many image tasks, with an ArcFace loss function \cite{deng2019arcface}. It reaches an accuracy of 99.78\% in the well-known LFW (Labeled Faces in the Wild) public benchmark for pair matching \cite{huang2008labeled}.
\textbf{LightCNN} was created to learn a compact embedding on large-scale face data with noisy labels. It has been reported to achieve state-of-the-art results on various face benchmarks without fine-tuning \cite{wu2018light}. In this work, we used the 29-layer model version, which reaches an accuracy of 99.40\% in LFW.

For evaluation, we used \texttt{face.evoLVe} \cite{wang2021face}, a face recognition library that provides a standard interface and can be used with various models for face-related analytics and applications. For the purposes of this research, the library was instrumented to keep track of individual errors. The instrumented library is available with our code release.

\subsection{Procedure}\label{subsec:procedure}

We performed an online user study, with the following structure. 

\subsubsection{Participant recruitment}
We recruited participants through a crowdsourcing platform for experimentation named Prolific.\footnote{\href{https://www.prolific.co/}{www.prolific.co}}
We considered four countries in continental Europe in which Prolific has large user bases: France, Germany, Italy, and Spain, plus the United Kingdom and Turkey. 
The crowdsourcing platform provides gender information and allows users to self-identify with a ``simplified ethnic group,'' which is made available as a criterion for participant selection.
We made sure that our sets of participants were gender balanced, and that for each pair of images, at least one person from each simplified ethnic group (White, Black, and Asian) participated in their evaluation. So, for every pair of images, we collected at least 3 annotations. For the subsequent analysis, for each pair of images, we take into account exactly 3 annotations, one from each simplified ethnic group.

In total, we recruited 235 participants, excluding 2 of them from our data due to failed attention checks. For the subsequent analysis, based on ethnic self-identification, we selected 162 participants.
Participants were paid 0.70 GBP to label 10 pairs of images, with an average completion time of 5 minutes. This amounts to 8.4 GBP per hour, which is slightly above the recommended payment by this platform (8 GBP/h).

\subsubsection{Demographic questionnaire}
\begin{figure}[p!]
\centering
    \begin{subfigure}[b]{0.9\textwidth}
        \begin{subfigure}[b]{0.475\textwidth}
             \centering
             \caption{Demographics questionnaire.}
             \includegraphics[width=\textwidth]{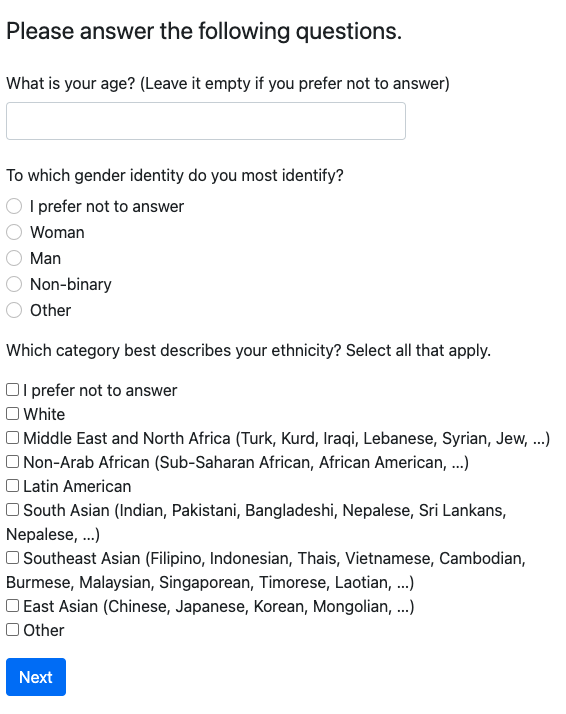}
             \label{fig:survey_demog}
         \end{subfigure}
         \begin{subfigure}[b]{0.475\textwidth}
             \centering
             \caption{Pair shown to the participant in survey. First question.}
             \includegraphics[width=\textwidth]{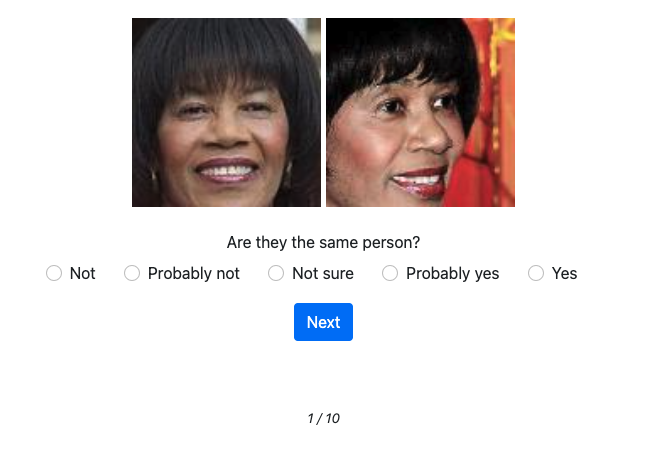}
             \label{fig:survey_pair}
        \end{subfigure}
    \end{subfigure}
    \hfill
    \begin{subfigure}[b]{0.9\textwidth}
          \centering
          \caption{Second question shown to the participant only when they answered something different to \textit{Yes} in the first question.}
         \includegraphics[width=\textwidth]{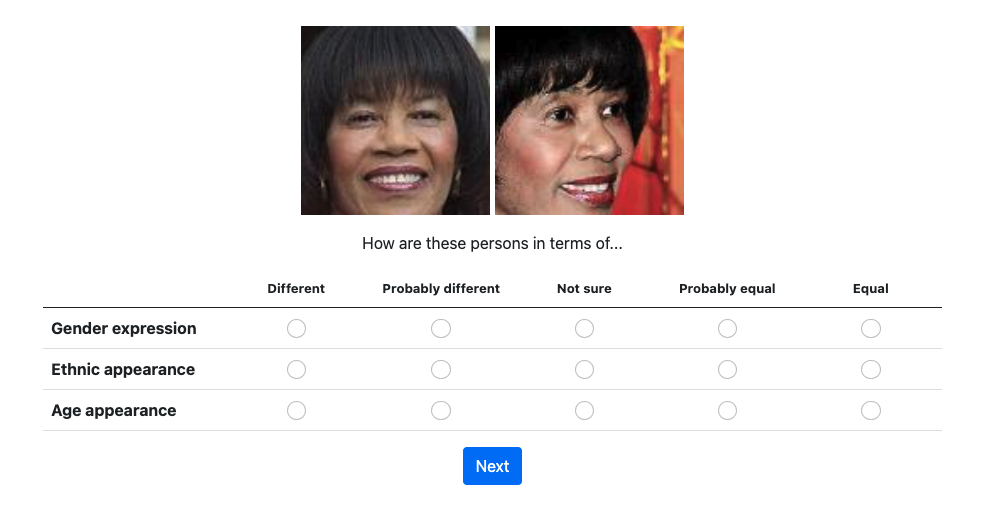}
         \label{fig:survey_clar}
    \end{subfigure}
    \caption{Survey screenshots. First, participants were asked about their age, gender identity and ethnic background. Then, participants started to evaluate the pairs of images. For those where the participant was not completely sure of both identities being the same person (answering something different to \textit{Yes} in question \ref{fig:survey_pair}), participants were asked to provide some details relate to gender expression and ethnic appearance similarities, as shown in \ref{fig:survey_clar}.}
    \label{fig:survey}
\end{figure}

Participants were asked about their age, gender identity, and ethnic background (see Figure \ref{fig:survey}).

\subsubsection{Face recognition tasks}
Participants evaluated one pair of images at a time. The participant had to answer the question \textit{Are they the same person?}, with the possible options: \textit{No}, \textit{Probably not}, \textit{Not sure}, \textit{Probably yes} or \textit{Yes}.

%
If the answer was different from \textit{Yes}, then the same pair of images was shown one more time, and the participant was asked about some of the differences between the two images.
These differences referred to gender expression, ethnic appearance, and age appearance (see Figure \ref{fig:survey_clar} for details). The participant had to answer three questions: \textit{How are these persons in terms of \{ gender expression | ethnic appearance | age appearance\}}. Each question had to be answered independently on a scale with five options: \textit{Different}, \textit{Probably different}, \textit{Not sure}, \textit{Probably equal} and \textit{Equal}.
We remark that we asked about ``expression'' and ``appearance'' because the participants do not know the identities of the photo subjects.

\subsubsection{Task selection} We found that the joint accuracy of the face recognition models (see \S\ref{subsec:measurements}) was correct above 95\% of the tasks.
Hence, due to budget constraints, we annotated all the cases where the models were wrong (``misses''), and a sample of cases in which both models were right (``hits'').
First, we annotated 363 ``misses'' (237 false negatives and 126 false positives, see Table \ref{tab:models_errors}), which were shown to a total of 164 participants, from which we selected a demographically balanced set of 108 participants. 
Next, we annotated 180 model ``hits,'' which were shown to a total of 69 participants, from which we selected a demographically balanced set of 54 participants. 
This selection of ``hits'' was a random sample that was demographically balanced for the true positive set (90 pairs) and for the true negative set (90 pairs). 

\subsection{Measurements}
\label{subsec:measurements}
We measured the following dependent variables.

\paragraph{Accuracy} Accuracy is defined as the fraction of correct responses with respect to the ground truth.

\begin{enumerate}
    \item \textbf{Machine accuracy}: Joint accuracy of the models. Each individual accuracy is calculated as the number of correct answers divided by the total number of pairs. To calculate the joint accuracy, in those cases where there is a disagreement between both models (for pairs labeled as positive by one model and as negative by the other) the average of their calibrated similarity scores is calculated and the label is decided based on this average (positive if it is above $0.5$ and negative if it is below).
    \item \textbf{Human accuracy}: Accuracy of the human annotators, as a group of three annotators. This is computed as a macro average, {\em i.e.}, first all the human evaluations on a pair of images are averaged, and then we determine whether that average is correct or not, computing human accuracy as number of correct responses by group of participants divided by the total number of pairs.
\end{enumerate}

\paragraph{Similarity} This is a measurement of how similar the model or the human annotator perceives the persons in the images.
\begin{enumerate}
    \item \textbf{ML similarity score}: Given two images, the model computes two embeddings or feature vectors (one per image). The numerical distance between these embeddings, $d$, is compared against a threshold $\theta$ to determine the output (if $d < \theta$, the pair of images is labeled as positive, while if $d > \theta$, the pair of images is labeled as negative). After normalizing this distance, we take $1-d$ as the similarity of the pair. Because the original scores are not calibrated, we calibrate this similarity, so it can be interpreted as a probability lying in the $[0, 1]$ interval. Scores close to 0.5 can be interpreted as a low model confidence.
    \item \textbf{Human perception of similarity}: this is inferred from the distance between the answer \textit{Not sure} and the annotator's actual answer to the questions specified in section \S\ref{subsec:procedure}. From this measurement we can infer human confidence: the answers in the extremes (\textit{No} and \textit{Yes}) correspond to the highest confidence, while answer \textit{Not sure} corresponds to the lowest confidence.
\end{enumerate}


\subsection{Ethical Considerations}

Our research plan was reviewed and approved by 
the Ethics Review Board of our university. 
The review included compliance with internationally accepted ethical principles in research, and with personal data protection guided by the EU General Data Protection Regulation (2016/679).

Regarding the gender and ethnicity discussed in this paper, it is important to note two things: (1) the original labels regarding gender and ethnicity in the testing database were inferred by means other than directly asking the person in the image about their demographics, so they should in no way be assumed to be true, and (2) in our study the participants were shown a pair of images and were asked about the similarity of \textit{gender expression} and \textit{ethnic appearance}, these being different concepts to those relating to the social identities of people in the images.

\section{Results}
\label{sec:results}
In what follows, we will consider a \textit{human error} when the mean response of the three annotators solving the same task corresponds to a wrong response, and a \textit{human success} when the mean response corresponds to a correct one. Equivalently, we will consider a \textit{machine error} when at least one of the two models solves the task incorrectly and a \textit{machine success} when both models are correct. For brevity, we will use "false positives", "false negatives", "true positives" and "true negatives" when we refer to the responses given by the models. In case we refer to the annotators' responses, we will do so explicitly ({\em e.g.}, Human False Negatives for positive pairs that annotators classified as negative). We will also show some significance test results (p-values, noted as $p$). Since what we want is to compare unknown a priori distributions, and by virtue of the continuity of our data, all these tests correspond to Kolmogorov-Smirnov tests.


\subsection{Participant Demographics}\label{subsec:participants}

Participants were on average 27.3 years old (SD=12.0 years). 
%
Out of the 82 participants that indicated their gender, 45 (55\%) identified as female, 34 (41\%) as male, and 3 (4\%) as non-binary.
The majority of the 205 participants that indicated an ethnicity identified as ``White'' (46\%), followed by ``Non-Arab African'' (19\%), ``South Asian`` (13\%), and ``East Asian'' (9\%). 
The remaining ethnicities accounted for less than 5\% of the participants each.

\subsection{Error Consistency (RQ1)}

\begin{figure}[t!]
\centering
    \begin{subfigure}[b]{0.32\textwidth}
         \centering
         \includegraphics[width=\textwidth]{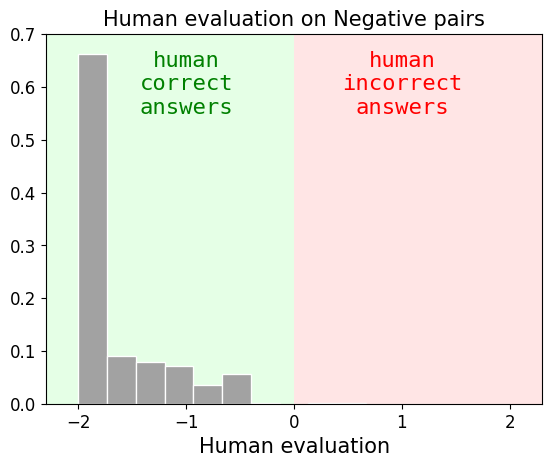}
         \caption{Different people}
         \label{fig:RQ1_neg}
    \end{subfigure}
    \hfill
    \begin{subfigure}[b]{0.32\textwidth}
        \centering
         \includegraphics[width=\textwidth]{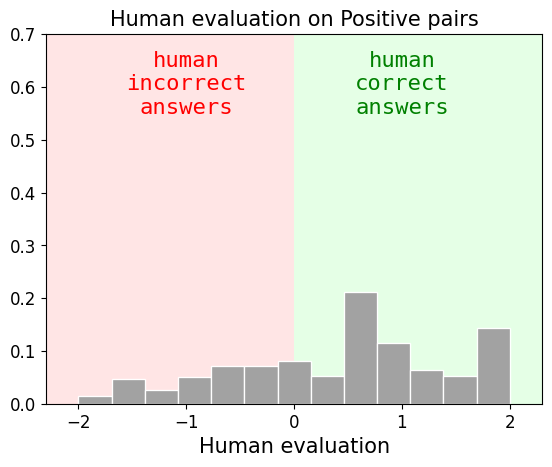}
         \caption{Same person}
         \label{fig:RQ1_pos}
    \end{subfigure}
    \hfill
    \begin{subfigure}[b]{0.32\textwidth}
          \centering
         \includegraphics[width=\textwidth]{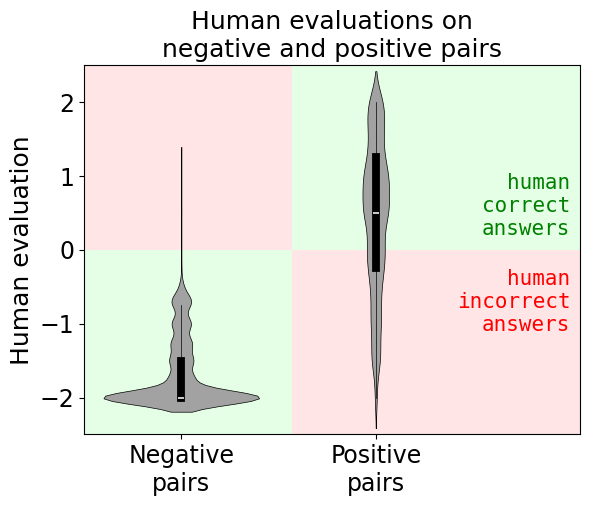}
         \caption{Comparison}
         \label{fig:RQ1_neg_and_pos_vio}
    \end{subfigure}
\caption{Human evaluations over 543 pairs of facial images. Negative pairs correspond of images of different people (216 negative pairs, see Figure \ref{fig:RQ1_neg}), while positive pairs correspond to images of the same person (327 negative pairs, see Figure \ref{fig:RQ1_pos}). In Figure \ref{fig:RQ1_neg_and_pos_vio}, we see the comparison of the distribution of negative pairs evaluations and positive pairs evaluations.
Responses range from -2 (``No'') to +2 (``Yes'').
}
\label{fig:RQ1}
\end{figure}

We now consider the agreement of human annotations, {\em i.e.}, the extent to which multiple people agree on whether a pair of images represents the same person or not.
Annotators were shown a total of 543 pairs of face images: 363 machine errors and 180 machine successes. Since the successes shown to the annotators are only a sample of all the successes from the models, we oversampled them to balance the workload. We also transformed every human annotation, originally based on a numeric 5-point scale, into a binary annotation in order to stablish a fair comparison between human and machine agreement. We obtained a \textit{moderate} multi-rater agreement among annotator responses (Fleiss' kappa = 0.47), which suggest that there is a mixture of agreement and disagreement between annotators (RQ1a). The moderate agreement present among the annotators is mainly due to the agreement they reach in those cases where the machine successes (Fleiss' kappa = 0.51), while in the cases where the machine makes a mistake we find no better agreement than would be the case by chance (Fleiss' kappa = -0.05).
As Figure \ref{fig:RQ1_neg} shows, human annotators are almost always correct in negative pairs, {\em i.e.}, when both images represent different people, as less than 5\% of pairs are incorrectly classified as positive by the annotators.
However, when images represent the same person, results are mixed. Figure \ref{fig:RQ1_pos} shows that although most of the positive pairs were correctly classified by the annotators, approximately 30\% of those pairs were incorrectly categorized as negative.
Differences in the distributions of labels on negative and positive pairs (as shown in the comparison of Figure~\ref{fig:RQ1_neg_and_pos_vio}) are significant at $p \ll 0.0001$.

For the models, we obtained an \textit{almost perfect} inter-rater agreement between the outputs of IR50 and LightCNN (Fleiss' kappa = 0.92), which suggests that the agreement between models is much better than would be expected by chance (RQ1b).
The human tendency to err with higher probability in positive pairs is similar to the models' way of erring: more than 65\% of model errors are false negatives (see Table~\ref{tab:models_errors}).
The agreement among human annotators becomes significantly lower when we consider only \textit{human errors} (Fleiss' kappa = $-0.05$), suggesting a poor agreement among annotators when their mean outcome is erroneous. This reduction in agreement is even more pronounced with the inter-rater agreement between models for \textit{machine errors} (Fleiss' kappa = $-0.29$), which suggests a great disagreement in tasks where at least one of the models made a mistake. The interpretation of negative values for Fleiss' kappa are based on \cite{landis1977measurement}.


\begin{table}[t]
\caption{Model errors, from a total of 5,460 evaluations. M1 stands for IR50, M2 stands for LightCNN, M1$\cap$M2 stands for the common cases, and M1$\cup$M2 stands for the union of cases with no repetition.}
    \centering
    \begin{tabular}{lcccc}
        \textbf{} & \textbf{ Model 1 } & \textbf{ Model 2 } & \textbf{ M1$\cap$M2 } & \textbf{ M1$\cup$M2 }\rule{0pt}{2.8ex}\\ \hline
        \textbf{ False Negatives } & 160 & 181 & 104 & 237 \rule{0pt}{2.8ex} \\ \hline
        \textbf{ False Positives } & 71 & 88 & 33 & 126 \rule{0pt}{2.8ex} \\ \hline
        \textbf{ Total } & 231 & 269 & 137 & 363 \rule{0pt}{2.8ex} \\ \hline
    \end{tabular}
\label{tab:models_errors}
\end{table}

\subsection{Error Alignment (RQ2)}




Next, we studied the extent to which human successes/errors are aligned with machine successes/errors.
%
%
We considered four categories of model outcomes: True Positives, False Negatives, True Negatives, and False Positives. 
Human performance when evaluating True Negative pairs was significantly different from human performance when evaluating False Positive pairs (\textit{p} $\ll$ 0.0001).
Differences were also significant in the case of human performance in the two subsets of positive pairs ($p \ll 0.0001$).

\begin{figure}[t!]
\centering
    \begin{subfigure}[b]{0.4\textwidth}
        \centering
        \captionsetup{width=.7\linewidth}
        \includegraphics[width=\textwidth]{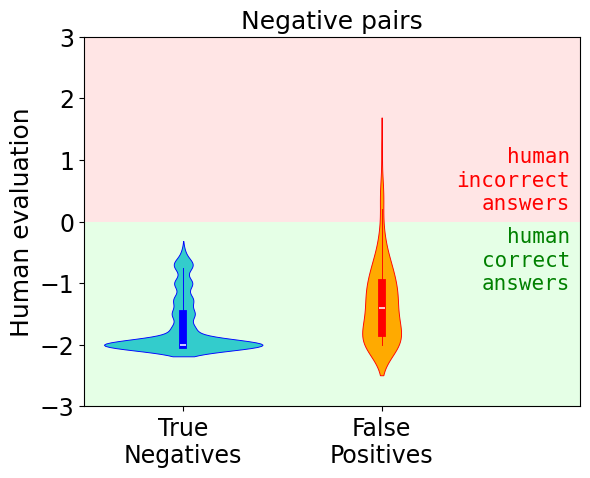}
        \caption{Different people}
        \label{fig:RQ2a_NEGVIO2}
    \end{subfigure}
    \begin{subfigure}[b]{0.4\textwidth}
          \centering
          \captionsetup{width=.7\linewidth}
         \includegraphics[width=\textwidth]{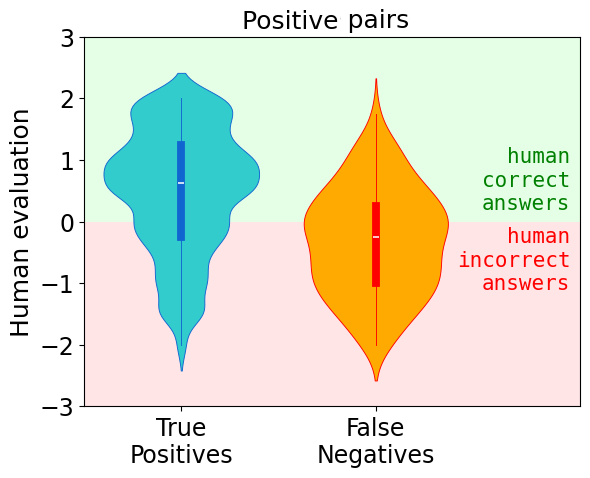}
         \caption{Same person}
         \label{fig:RQ2a_POSVIO2}
    \end{subfigure}
\caption{Human evaluations of machine errors (in orange, right violin in every figure) and successes (in blue, left violin in every figure). Negative pairs (a) correspond to images of different people, with False Positives indicating pairs that the models mistakenly labeled as the same person.
Positive pairs (b) correspond to images of the same person, with False Negatives indicating pairs that the models mistakenly labeled as different people.
Responses range from -2 (``No'') to +2 (``Yes'').}
\label{fig:RQ2a2}
\end{figure}

In the case of negative pairs, in which human annotators are almost always correct, Figure~\ref{fig:RQ2a_NEGVIO2} shows less certainty and a possibility of error in the pairs in which ML models make a mistake.
Human annotators are less likely to select the option ``No'' and more likely to select the option ``Probably not'' when asked about a pair of images of different people for which the ML models mistakenly indicated that they were the same person. 
In the case of positive pairs, shown in Figure~\ref{fig:RQ2a_POSVIO2}, we see a similar trend. In this situation, human errors are concentrated in the cases in which the models also made an error. In other words, there are some pairs of images of the same person for which both human annotators and models are likely to err.
This, put together with the significance above, suggests that humans find cases where the machine erred more difficult in comparison to those where the machine succeed (RQ2a).

\begin{figure}[t!]
\centering
    \begin{subfigure}[b]{0.4\textwidth}
         \centering
         \includegraphics[width=\textwidth]{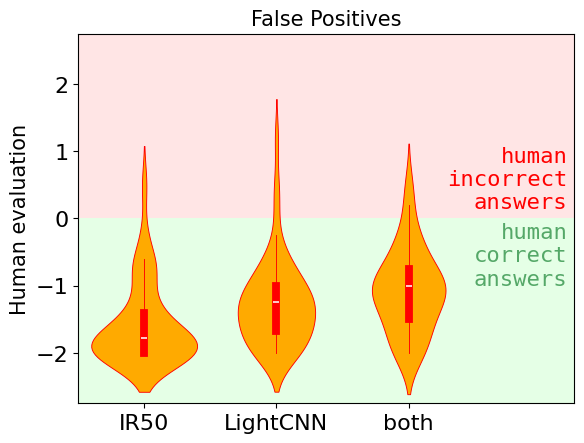}
         \caption{}
         \label{fig:RQ2b_FP}
    \end{subfigure}
    \begin{subfigure}[b]{0.4\textwidth}
          \centering
         \includegraphics[width=\textwidth]{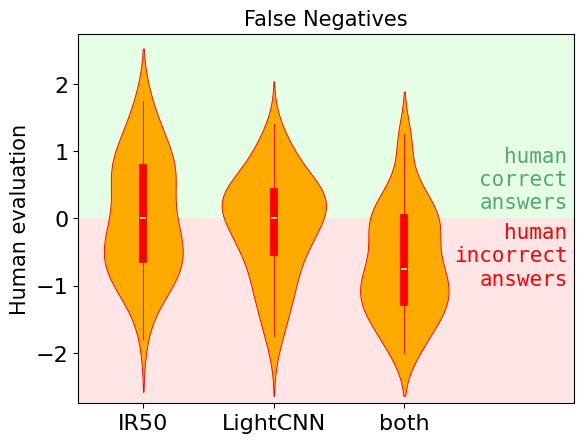}
         \caption{}
         \label{fig:RQ2b_FN}
    \end{subfigure}
\caption{Human evaluations of machine errors made by model IR50 only, errors made by model LightCNN only, and errors made by both models. Responses range from -2 (``No'') to +2 (``Yes'').}
\label{fig:RQ2b}
\end{figure}

%
In general, annotators were more likely to make a mistake on pairs in which both models made an error (RQ2b); with human certainty (preference for ``No'' over ``Probably not'') reduced in false positives of both models, and human error more likely in false negatives of both models.
%
In the case of False Positives, human evaluation over those errors committed solely by IR50 are significantly different from human evaluations over those committed by both models, at $p<0.001$. 
However, human evaluations over False Positives committed solely by LightCNN is not significantly different from human evaluation over False Positives committed by both models ($p=0.17$). 
We depict these differences in Figure \ref{fig:RQ2b_FP}.
In the case of False Negatives, human evaluation over those committed solely by IR50 is significantly different from human evaluation over those committed by both models ($p<0.001$). 
Similarly, human evaluations over False Negatives committed solely by LightCNN are significantly different from human evaluation over False Negatives committed by both models ($p \ll 0.0001$). 
We depict these differences in Figure \ref{fig:RQ2b_FN}.

\begin{figure}[t!]
\centering
    \includegraphics[width=0.95\textwidth]{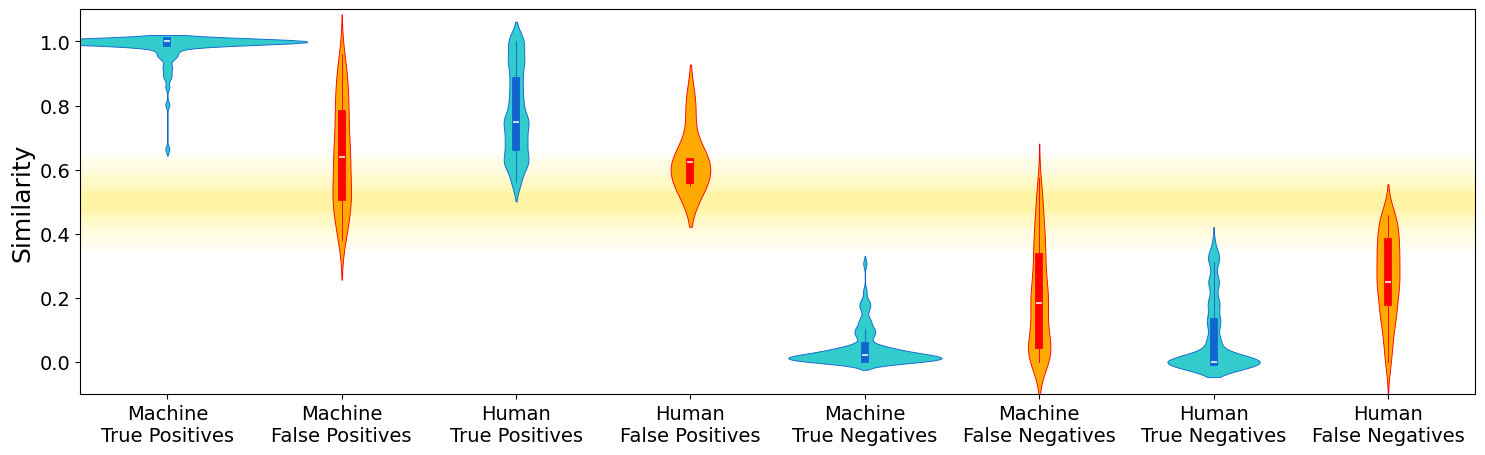}
    \caption{Human annotators perception of similarity and machine similarity score for different categories of human/machine errors (in orange) and successes (in blue). Note that machine confidence can be inferred from the similarity score (the further the similarity is from 50\%, the higher the confidence). The yellow band near a similarity score of 0.5 includes machine errors that can be anticipated as possible errors.}
    \label{fig:RQ2c}
\end{figure}

We examined human annotators' perception of similarity and compared them with model-computed similarity scores.
This time we distinguished between eight overlapping categories of human and model errors and successes: \{ Human, Machine \} $\times$ \{ True Positives, False Positives, True Negatives, False Negatives \}.
When both models and annotators gave correct responses, there were differences between the machine similarity score and annotator's perception of similarity (see blue violins in Figure \ref{fig:RQ2c}).
We found significant differences between both similarities for positive cases ($p \ll 0.0001$), and for negative cases ($p \ll 0.0001$).

This analysis reveals differences in the distribution of machine similarities, which tend to be bimodal and concentrated on the extremes, while human perceptions of similarity are more nuanced and dispersed (RQ2c). 
We found significant differences when both models and annotators gave incorrect responses (see orange violins in Figure \ref{fig:RQ2c}) for negative pairs ($p \ll 0.0001$), 
but not for positive pairs ($p=0.35$).
In the case of False Positives, annotators' perception of similarity when claiming a negative pair as positive tended to accumulate close to 0.5, indicating a low confidence in their answers (for comparison with machine similarity scores, human similarity 0.5 corresponds to the case ``Not sure'').
The yellow band around similarity 0.5 in Figure \ref{fig:RQ2c} includes machine errors that based on these observations could be predicted in advance as potential errors.

\subsection{The Role of Gender and Ethnicity (RQ3)}


\begin{figure}[t!]
\centering
        \centering
        \includegraphics[width=0.4\textwidth]{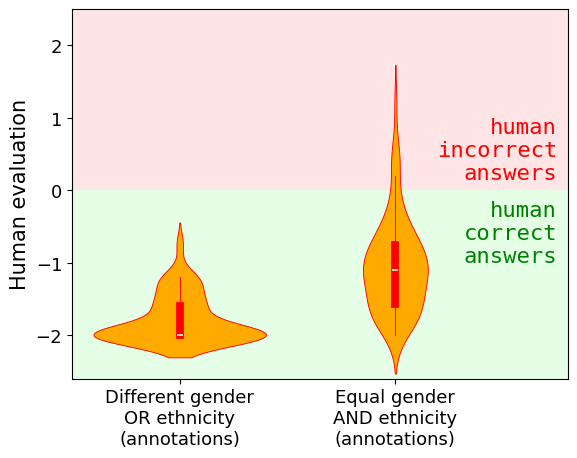}
         \caption{Human evaluation over False Positive machine errors. We examined pairs of images annotated as having different gender expression \textbf{or} different ethnicity appearance, compared to pairs annotated as having similar gender expression \textbf{and} similar ethnicity appearance. This indicates that annotators are less confident in differentiating between two distinct identities when they observe similarities in terms of gender expression and ethnic appearance.}
         \label{fig:RQ3a_annotations}
\end{figure}

We examined human evaluations over two categories of False Positive errors, {\em i.e}, cases of the different people mistakenly identified by a model as being the same person.
We compared pairs of facial images annotated as different in terms of gender expression \textbf{or} ethnic appearance, against pairs of facial images annotated as similar in gender expression \textbf{and} similar ethnic appearance.
Figure \ref{fig:RQ3a_annotations} shows the results, and indicates that humans are less certain about their answer to the question on whether both images depict the same person ({\em i.e.}, more likely to indicate ``Probably not'' and less likely to indicate ``No'') for those pairs annotated as having equal or similar gender expression and similar or equal ethnicity appearance (RQ3a).
Differences are significant at $p \ll 0.0001$.
%

\begin{figure}[t]
\centering
    \includegraphics[width=0.85\textwidth]{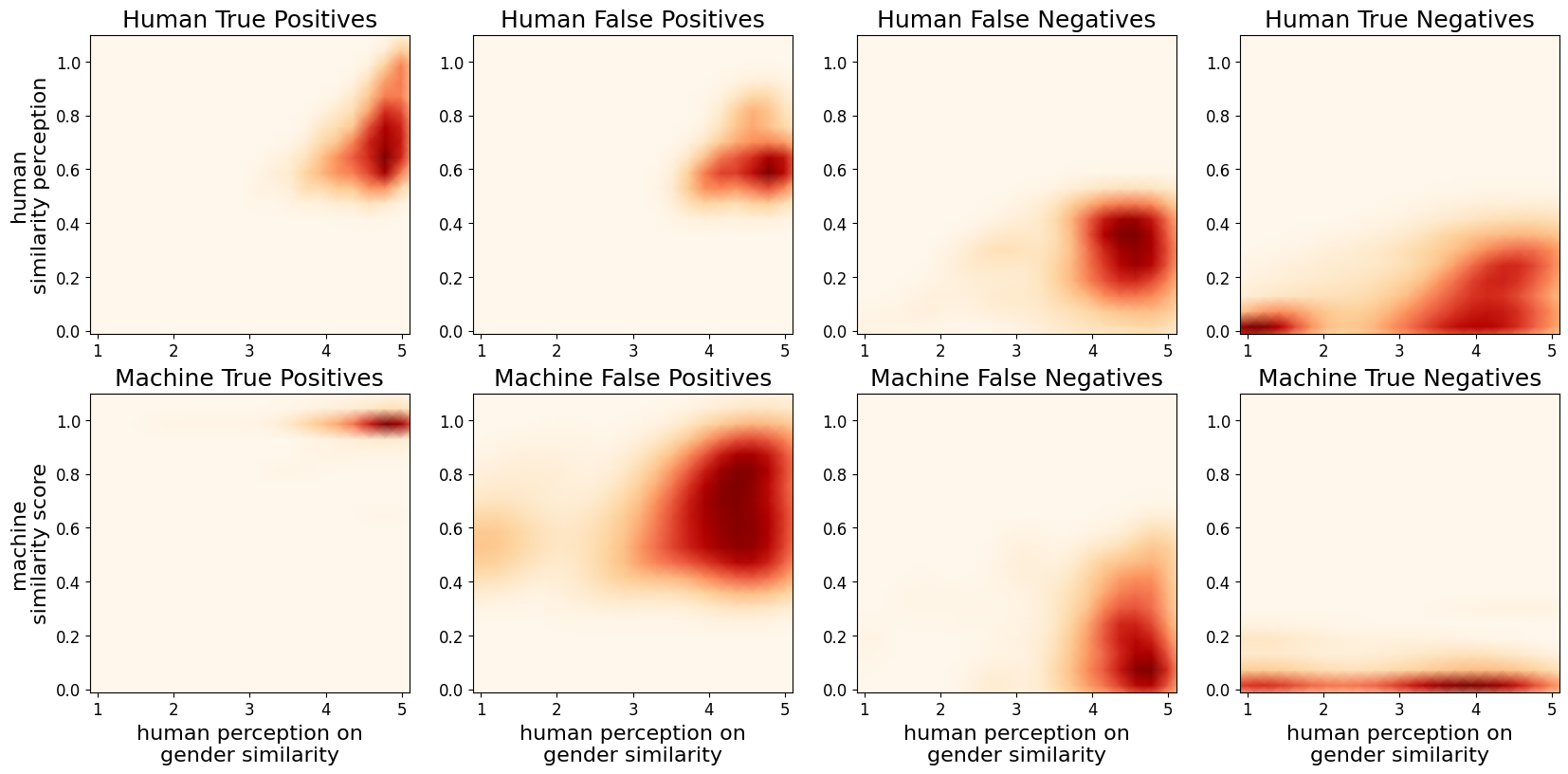}
    \caption{Distribution of human perception of gender expression similarity (1 - Different, 5 - Equal) for different human and machine outcomes, compared to human similarity perception, and machine similarity score, respectively.}
    \label{fig:RQ3c_gender}
\end{figure}
\begin{figure}[t]
\centering
\includegraphics[width=0.85\textwidth]{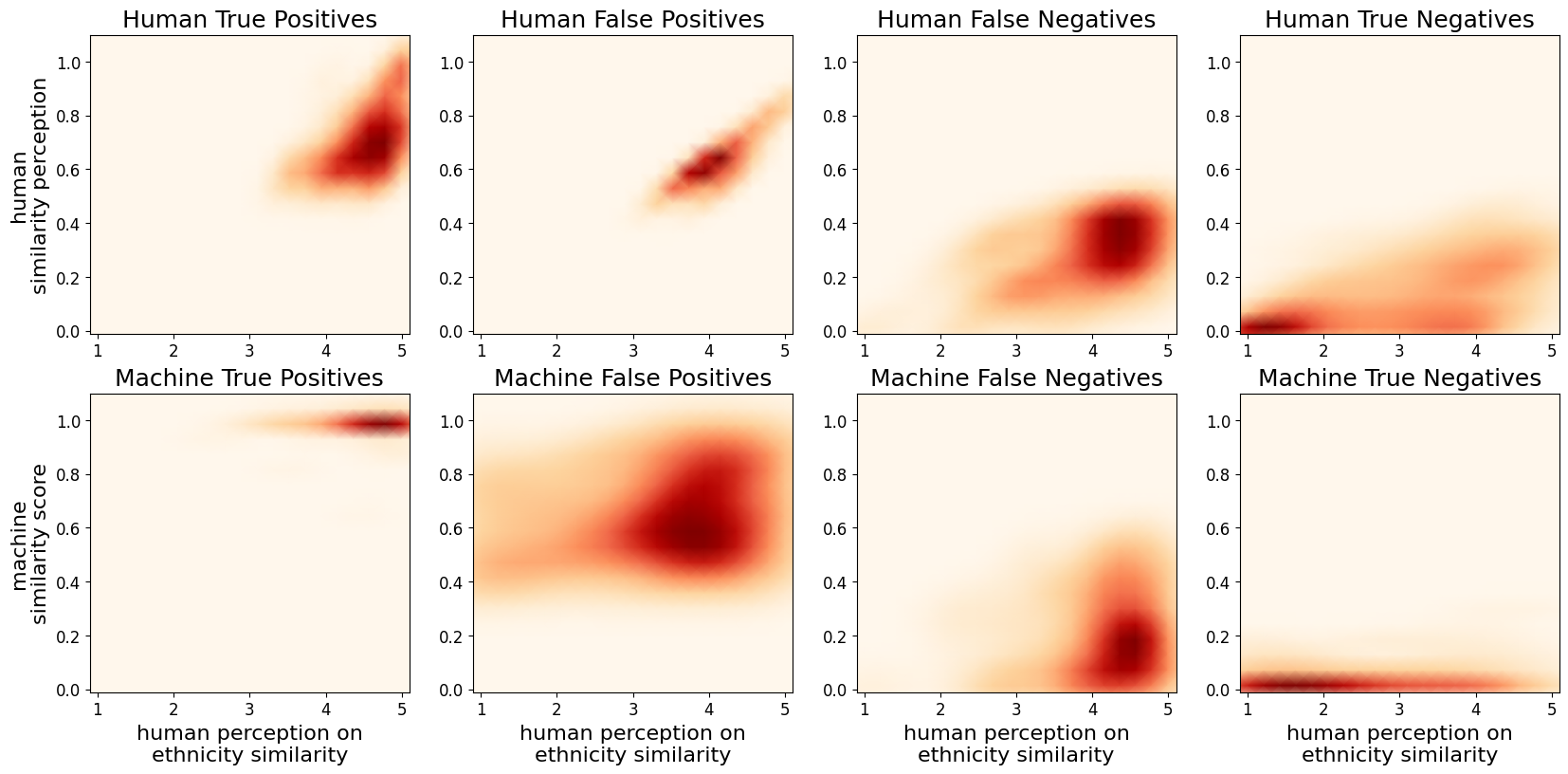}
    \caption{Distribution of human perception of ethnicity appearance similarity (1 - Different, 5 - Equal) for different human and machine outcomes, compared to human similarity perception and machine similarity score, respectively.}
    \label{fig:RQ3c_ethnicity}
\end{figure}

Figure~\ref{fig:RQ3c_gender} depicts two-dimensional plots in which we compare the human perception of gender expression similarity (in the $x$ axis) against the human perception of similarity of the images (in the $y$ axis).
For the human perception of ethnicity similarity, the visual patterns are very similar (see Figure \ref{fig:RQ3c_ethnicity}).
%
%
The user interface for this question is the one showed in Figure \ref{fig:survey}.

%
People state that images of the same person portray the same gender expression and ethnic appearance.
This is more evident (less noisy) in the case of gender expression, suggesting that this signal determines more directly the human label than ethnic appearance. Another possible explanation is that ethnic appearance might be more affected by different lightning conditions in the images (RQ3b).

In our experiments, we found partial evidence of the ``other-race'' effect (see Table \ref{tab:other-race}). We calculated the error rates for the three self-ascribed ethnicities: White, 
Black, 
and Asian. 
We considered only pairs of images with the same ethnicity label in both images and computed the error rate for each of these sets of pairs.
``White'' annotators are the most accurate when annotating images of ``White'' people, and ``Black'' annotators are the most accurate when annotating images of ``Black'' people, but this was not the case for ``Asians''.

\begin{table}[t]
\caption{Human and Machine error rate. First three rows are demographic groups evaluating different set of pairs. "White-white" pairs stands for pairs containing images of two people labeled as white, and so forth.}
\centering
\setlength{\tabcolsep}{10pt}
\begin{tabular}{@{}lccc@{}}
\toprule
& white-white pairs  & black-black pairs & asian-asian pairs \\ \midrule
White & 0.10 & 0.45 & 0.20 \\ 
Black & 0.55 & 0.04 & 0.02 \\
Asian & 0.18 & 0.06 & 0.09 \\
Machine & 0.09 & 0.07 & 0.09 \\ \bottomrule
\end{tabular}
\label{tab:other-race}
\end{table}

\subsection{Exploratory study of error-based human-machine collaboration (RQ4)}

We conducted a study with the intention of illustrating the consequences of applying a human supervision strategy based on the results previously obtained. We studied the improvement over model accuracy that would result from manually reviewing the pairs evaluated by the machine. As explained in the experimental settings, the ``machine accuracy'' is the accuracy resulting from the combined performance between the IR50 and LightCNN models. Under these considerations, the accuracy achieved by both models jointly is 93.5\%. 

\begin{figure}[t!]
\centering
    \begin{subfigure}[b]{0.45\textwidth}
         \centering
         \includegraphics[width=\textwidth]{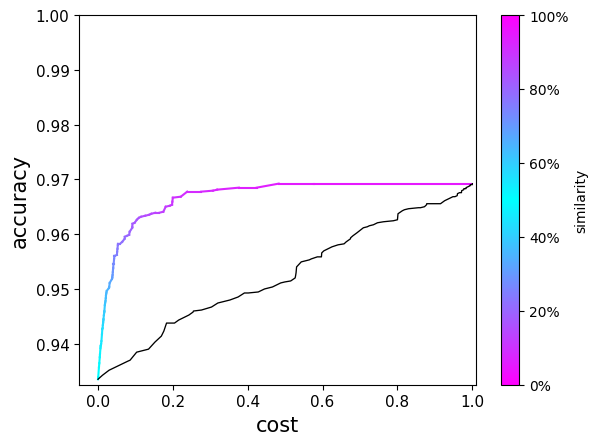}
         \caption{}
         \label{fig:pilot_similarity}
    \end{subfigure}
    \begin{subfigure}[b]{0.45\textwidth}
         \centering
         \includegraphics[width=\textwidth]{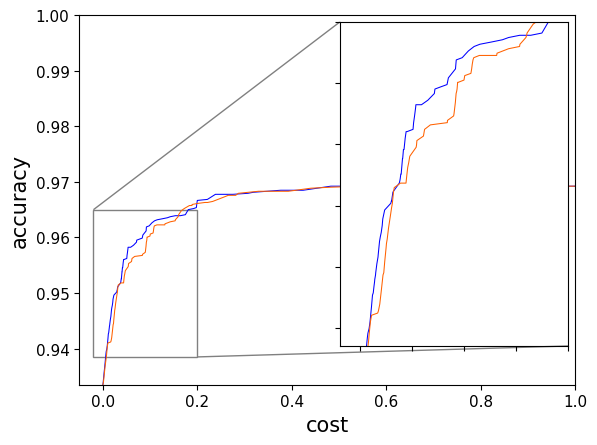}
         \caption{}
         \label{fig:pilot_similarity_both}
    \end{subfigure}
    \caption{Human evaluation of pairs classified by the machine. On the left, (a) shows the evolution of joint human-machine accuracy when annotators evaluate pairs in increasing order of machine certainty, inferred from machine similarity score. On the right, (b) shows the evolution of joint human-machine accuracy if, in addition to the previous strategy, we prioritized those pairs where the models gave different answers (blue line). This accuracy exceeds the joint accuracy obtained when this priority is not taken into account (orange line).
    The initial machine accuracy was 93.5\%. Cost represents the rate of the number of annotators.}
    \label{fig:pilot_similarity_}
\end{figure}

The first improvement is based on  the results obtained related to RQ2c: the use of machine confidence to prioritize those cases that have a high probability of being corrected by the human annotator. Note that machine confidence can be inferred from the similarity score (the further the similarity is from 50\%, the higher the confidence, see Figure \ref{fig:RQ2c}). The evolution of joint accuracy when this prioritization is implemented can be seen in the colored line in Figure \ref{fig:pilot_similarity}. We can observe that the pairs that human annotators are able to solve correctly are concentrated at the beginning of the workflow, leading to an early and rapid growth of the joint accuracy. This marked improvement in accuracy contrasts with the results we would obtain if this strategy were not taken into account (see the black line in Figure \ref{fig:pilot_similarity}).

The second improvement is based on the results obtained when investigating RQ2b: prioritizing those pairs where the models gave different answers, {\em i.e.}, only one of the two models correctly classified the pair. As we can see in Figure \ref{fig:pilot_similarity_both}, the joint accuracy obtained if these pairs are prioritized (blue line) exceeds the joint accuracy obtained when this priority is not implemented (orange line) during most of the human annotation flow, especially at the beginning.

\section{Discussion}
\label{sec:discussion}
This study comparing human errors and machine errors allows us to develop strategies for enhancing the efficiency and effectiveness of human-computer collaboration in the domain of face recognition tasks. 
In particular, we should try to make the most of human capabilities to complement machine deficits, and viceversa.
%
Three main observations can be deduced from the results obtained by investigating the first three research questions posed at the beginning of this paper.

First, sufficient consistency was observed in both the annotations provided by the humans and the responses generated by the model. The consistency in human responses enables us to identify ``non-human-like" errors, which are errors that models make but are unlikely to be made by humans.
%
In our setting, these are false positive pairs.
When examining the correlation between human errors and machine errors, it was observed that when the machine makes a mistake, humans are more likely to make a mistake as well, particularly in the case of positive pairs, where humans have a high error rate in comparison with the machine false negative error rate. However, in negative pairs, although human confidence decreases for those cases where the machine fails, human responses are mostly accurate. While humans encounter more challenges with machine false negatives compared to true positives, humans seem to find no challenge with machine false positives, suggesting that the machine struggles with certain negative pairs, while humans do not find them difficult.

Next, when we categorize machine errors into those occurring in just one of the two models and those occurring in both models, we observe two distinct patterns for positive and negative pairs when compared to human assessments. In the case of a machine false negative pair, humans are notably more prone to error when that false negative is committed by both models. This reveals a correlation between the challenges faced by both models and those faced by humans when assessing positive pairs. However, when a human assesses a machine false positive pair, the probability of error is not significantly influenced by whether the error is common to both models or not, which is consistent with the observation above. However, what is significantly influenced is the change in human certainty, which is smaller, but still correct.

These two observations indicate that (1) humans have a significantly better capacity to distinguish negative pairs compared to machines. Of the 126 false positives made by the machine, humans made only 6 errors (4.8\%), and successfully identified all negative pairs correctly classified by the models. And (2) humans have a significantly better capacity to correctly classify those pairs in which both models disagree, over those pairs in which both models are wrong.

Finally, we observed that only when humans and machine make an error by failing to detect a positive pair, their similarity scores could become similar. In all other scenarios of correct and incorrect identifications, the human and the machine provide responses based on notably different similarity scores. Moreover, there is a substantial disparity between the machine's similarity rating for a correct identification and the machine's similarity score for an incorrect one. This difference is even more pronounced when the machine classifies a pair as positive. 
This, combined with the high accuracy mentioned above of humans over machines in evaluating certain pairs, could help to anticipate potential errors and suggest that a manual examination of these cases by a group of annotators could be beneficial.

Based on these observations, following the suggested strategies we could improve the accuracy of the system by approximately 3 percentage points by assuming 10\% of the total cost only ({\em i.e.}, by assuming the cost of 546 human annotations). This improvement of 3\% is equivalent in our case to the correction of 148 pairs (98 negative and 50 positive) misclassified by the machine, which outweighs the improvement of just 0.4\% (equivalent to 27 corrections, 19 negative and 8 positive pairs) that we would achieve if we did not follow the proposed strategies. This study is a clear example of how the findings of such comparative and exploratory analysis can facilitate the proposal of simple but powerful human-machine interaction paradigms.

\subsection{Limitations and future work}
Approaching more real-world use cases also highlights a possible limitation of the work developed here. In use cases for face recognition technologies, the nature of the domain determines under which thresholds of similarity score (and, therefore, machine confidence) the machine's response is considered positive or negative. In cases where, for example, it is desirable to prioritize the reduction of false positives without the possible increase of false negatives being detrimental ({\em e.g.}, face recognition methods for private access controls \cite{ibrahim2011study}), the similarity score is set at a higher value than in other scenarios where it is desirable to prioritize the reduction of false negatives ({\it e.g.}, face recognition methods for law enforcement \cite{raposo2023use}). This work has been approached from a symmetric costs point of view (with a threshold for the similarity score of 50\%) and thus serves as an example or starting point for possible different scenarios to be readjusted.

Also, the error behavior of a system (and beyond errors, its overall behavior) is highly dependent on the training data and the architecture of the chosen model. In this work we have chosen two specific pre-trained models (IR-50 \cite{he2016deep} and LightCNN \cite{wu2018light}), well known in the literature, taking care that the training was based on the same dataset MS-Celeb-1M \cite{guo2016ms} for a fair comparison. In this sense, RQ1a and RQ1b in \S\ref{subsec:consistency} could be interpreted, rather than as research questions, as prerequisites. The consistency alluded in RQ1 allows us to propose an error characterization, a differentiating axis between human errors and machine errors, on which the exploration and comparisons developed in this work are then based. It is therefore important to bear in mind that in other different scenarios this condition might not be present.

A possible future work is based on revisiting some of the biases that may occur in a human-machine interaction scenario taking into account the results of this analysis. Biases such as algorithmic aversion, overconfidence, or confirmation bias can vary significantly depending on whether the resolution offered by the machine is more or less similar to the resolution that a human agent could offer. More specifically, our results suggest that machine aversion is more likely to be found in scenarios where minimizing false negatives is prioritized, as this will increase the proportion of false positives and, as this is a rare error in humans, may cause more rejection.

\section{Conclusions}
\label{sec:conclusions}
The main conclusions drawn from this work are the following: 
\begin{enumerate}
    \item The facial recognition models shows a marked disparity in similarity scores between correctly and incorrectly resolved pairs.
    \item There is a correlation between the shared challenges faced by the models (errors made by both models and not just one) and the difficulties experienced by humans, whereas humans encounter fewer issues when classifying pairs where the models provided different results.
    \item Humans perform substantially better than facial recognition models in assessing negative pairs (pairs consisting of different identities).
\end{enumerate}

Observation (1) enabled us to detect potential errors in the facial recognition models, while observations (2) and (3) helped us prioritize those potential errors that a human annotator has a high chance of correcting. Implementing this in practice allowed us to design a manual evaluation strategy that achieves maximum joint human-machine precision with a very low number of annotations.

In addition to the quantitative improvements shown in this work, it is worth paying attention to the impact that some of these conclusions could have on facial recognition tasks in real-world contexts.
As we saw in the human-machine collaboration paradigm proposed above, most of the machine errors corrected by a human annotator are negative pairs that were predicted as positive by the model. 
%
It is worth noting that this should not only be taken into account when a face recognition system is already in the development or deployment phase, but also when evaluating the suitability of integrating an automatic face recognition system in the specific application domain.
In scenarios where the occurrence of false positives might have serious consequences and potentially affect fundamental rights, if there is a concern of lack of adequate human oversight, the integration of facial recognition technologies demands a rigorous and thoughtful reconsideration.

In use cases where the resolution of face recognition tasks by a machine learning system can be conveniently monitored by human reviewers, it is still imperative to implement oversight strategies that acknowledge and address the disparate error patterns exhibited by humans and machines. 

A noteworthy aspect is the observation that the human advantage over machines in assessing negative pairs might be linked to the perception that humans have built on notions of similarity and difference in gender expression and ethnic appearance. 
Given a pair of images corresponding to two different identities, if a human makes the mistake of saying that they are the same person (which, as we have seen, happens infrequently), it does so in the belief that both identities share a similar gender expression and ethnic appearance. 
When the human correctly classifies a negative pair that was classified as positive by the machine, both perceptions of gender and ethnicity seem to play a distinctive role in the final human decision.
This apparent human tendency to use gender and ethnicity-related characteristics to differentiate negative pairs could be due not only to gender and racial stereotypes perpetuated in society, but also to the predominant presence of stereotypical images in face recognition databases \cite{dominguez2022gender, keyes2018misgendering}.

Finally, our results suggest that facial recognition algorithms are not advanced enough to fully replace human roles in real world scenarios. This may also not be desirable, especially in light of the recent ethical and legal concerns that have been raised about the use of this technology. The current draft of the EU AI Act \cite{EUAIAct} contains many explicit and implicit allusions to facial processing, whose applications are considered at different risk levels, including \textit{high risk} and \textit{forbidden}. This envisions a future scenario for face recognition technologies in which permanent human oversight will be essential, highlighting the value of preserving human input in decision-making.

\section*{Acknowledgement}
MCIN/AEI/10.13039/501100011033 under the Maria de Maeztu Units of Excellence Programme (CEX2021-001195-M).




\bibliographystyle{elsarticle-num} 
\bibliography{mybib.bib}






\end{document}